\documentclass[
    ,final            % use final for the camera ready runs
  ]
  {aipproc}

\layoutstyle{6x9}

\newcommand{\be}{\begin{equation}}
\newcommand{\ee}{\end{equation}}
\newcommand{\bea}{\begin{eqnarray}}
\newcommand{\eea}{\end{eqnarray}}

\newcommand{\xbj}{x_{\!\scriptscriptstyle B}}

\newcommand{\bfk}{\mbox{\boldmath $k$}}

\newcommand{\pup}{p^\uparrow}

\newcommand{\bfp}{\mbox{\boldmath $p$}}

\def\lsim{\mathrel{\rlap{\lower4pt\hbox{\hskip1pt$\sim$}}\raise1pt\hbox{$<$}}}
\def\gsim{\mathrel{\rlap{\lower4pt\hbox{\hskip1pt$\sim$}}\raise1pt\hbox{$>$}}}
\def\nostrocostruttino#1\over#2{\mathrel{\mathop{\kern 0pt \rlap
{\hbox{$#1$}}} \hbox{\kern-.135em $#2$}}}

\newcommand{\NP}[1]{{\it Nucl.\ Phys.}\ {\bf #1}}
\newcommand{\ZP}[1]{{\it Z.\ Phys.}\ {\bf #1}}
\newcommand{\PL}[1]{{\it Phys.\ Lett.}\ {\bf #1}}
\newcommand{\PR}[1]{{\it Phys.\ Rev.}\ {\bf #1}}
\newcommand{\PRL}[1]{{\it Phys.\ Rev.\ Lett.}\ {\bf #1}}

\def\kt{k_\perp}

\def\ptq{p_\perp}
\def\pt{P_T}

\begin{document}

\title{Understanding the role of Cahn and Sivers effects in Deep Inelastic Scattering\footnote{Talk delivered by A. Prokudin at  DIS05, Madison, WI, USA}}


\author{M.~Anselmino}{
  address={Dipartimento di Fisica Teorica, Universit\`a di Torino and \\
          INFN, Sezione di Torino, Via P. Giuria 1, I-10125 Torino, Italy}
}

\author{M.~Boglione}{
  address={Dipartimento di Fisica Teorica, Universit\`a di Torino and \\
          INFN, Sezione di Torino, Via P. Giuria 1, I-10125 Torino, Italy}
}

\author{U.~D'Alesio}{
  address={INFN, Sezione di Cagliari and Dipartimento di Fisica,  
Universit\`a di Cagliari,\\
C.P. 170, I-09042 Monserrato (CA), Italy}
}
\author{A. Kotzinian}{
  address={Dipartimento di Fisica Generale, Universit\`a di Torino and \\
          INFN, Sezione di Torino, Via P. Giuria 1, I-10125 Torino, Italy}
  ,altaddress={Yerevan Physics Institute, Alikhanian Brothers St. 2; AM-375036
          Yerevan, Armenia; \\
          JINR, Dubna, 141980, Russia} % additional visiting address
}
\author{F.~Murgia}{
  address={INFN, Sezione di Cagliari and Dipartimento di Fisica,  
Universit\`a di Cagliari,\\
C.P. 170, I-09042 Monserrato (CA), Italy}
}

\author{A.~Prokudin}{
  address={Dipartimento di Fisica Teorica, Universit\`a di Torino and \\
          INFN, Sezione di Torino, Via P. Giuria 1, I-10125 Torino, Italy}
}

\begin{abstract}
The role of intrinsic $\bfk_\perp$ in semi-inclusive Deep
Inelastic Scattering (SIDIS) processes ($\ell \, p \to \ell \, h \, X$) is studied
with exact kinematics within QCD parton model at leading order; the dependence
of the unpolarized cross section on the azimuthal angle between the leptonic
and the hadron production planes (Cahn effect) is compared with data and used
to estimate the average values of $k_\perp$ both in quark distribution and
fragmentation functions. The resulting picture is applied to the description
of the weighted single spin asymmetry $A_{UT}^{\sin(\phi_\pi - \phi_S)}$
recently measured by the HERMES collaboration at DESY; this allows to extract
parameters for the quark Sivers functions. The extracted Sivers functions give
predictions for the COMPASS measurement of $A_{UT}^{\sin(\phi_\pi - \phi_S)}$
in agreement with recent data, while their contribution to HERMES
$A_{UL}^{\sin\phi_\pi}$ is computed and found to be small. Predictions for
$A_{UT}^{\sin(\phi_K - \phi_S)}$ for kaon production at HERMES are also given.
\end{abstract}

\maketitle

%%%%%%%%%%%%%%%%%%%%%%%%%%%%%%%%%%%%%%%%%%%%
%% MAINMATTER
%%%%%%%%%%%%%%%%%%%%%%%%%%%%%%%%%%%%%%%%%%%%
The role of intrinsic $\bfk_\perp$ is known to be important 
in unpolarized SIDIS
processes~\cite{cahn} and becomes crucial for the explanation of many single 
spin effects recently observed and still under active investigation in several 
ongoing experiments; spin and $\bfk_\perp$ dependences can couple in parton 
distribution and fragmentation functions,
%\cite{aram}$^,$\cite{rev}
giving 
origin to unexpected effects in polarization observables. One such example 
is the azimuthal asymmetry observed in the scattering of unpolarized leptons 
off polarized protons~\cite{hermUL},~\cite{hermUT} and 
deuterons~\cite{compUT}. 

A recent analysis of Single Spin Asymmetries (SSA) in $\pup \, p \to \pi \, X$ processes, with a 
separate study of the Sivers and the Collins contributions, has been performed 
respectively in Refs.~\cite{fu} and~\cite{noicol}, with the conclusion
that the Sivers~\cite{siv} mechanism alone can explain the data~\cite{e704}, 
while the Collins~\cite{col} mechanism is strongly suppressed.   

We considered~\cite{our} the role of parton intrinsic motion in SIDIS processes 
within the QCD parton model at leading order. The average values of $k_\perp$ for 
quarks inside protons, and $\ptq$ for final hadrons inside the fragmenting 
quark jet, are fixed by comparison with data~\cite{cahndata} on the dependence 
of the unpolarized cross section on the azimuthal angle between the leptonic 
and the hadronic planes and on $P_T$. 

Within the factorization scheme, assuming an independent fragmentation 
process, the SIDIS cross section for the production of a hadron $h$ in 
the current fragmentation region with the inclusion of all intrinsic 
motions can be written as~\cite{our}
\bea
\frac{d^5\sigma^{\ell p \to \ell h X }}{d\xbj \, dQ^2 \, dz_h \, d^2 {\bf P}_T} 
= \sum_q  e_q^2 \int  \! d^2 {\bf k_\perp} \; f_q(x,\bfk _\perp) \; 
\frac{2\pi\alpha ^2}{\xbj ^2 s^2}\,\frac{\hat s^2+\hat u^2}{Q^4}\; 
\label{sidis-Xsec-final} \\  
\times D_q^h(z,\bfp _\perp) \; \frac{z}{z_h} \, 
\frac{\xbj}{x}\left( 1 + \frac{\xbj^2}{x^2}\frac{\kt^2}{Q^2} \right)^{\!\!-1} 
\> \cdot  \nonumber
\eea
%Details can be found in Ref. \refcite{our}. 
It is instructive, and often 
quite accurate, to consider the above equation
in the much simpler limit in which only terms of ${O}(\kt/Q)$ are 
retained. In such a case $x \simeq \xbj, z \simeq z_h$ and 
${\bf p_\perp} \simeq {\bf P}_T - z_h \, {\bf k_\perp}$.
In what follows we assume, both for parton densities and fragmentation 
functions, a factorized Gaussian $k_\perp$  and $p_\perp $ dependence.
%$
%f_q(x,\bfk_\perp) = f_q(x) \, \frac{1}{\pi \langle\kt^2\rangle} \,
%e^{-{\kt^2}/{\langle\kt^2\rangle}}
%$
%and 
%$
%D_q^h(z,\bfp _\perp) = D_q^h(z) \, \frac{1}{\pi \langle{p_\perp}^2\rangle}
%\, e^{-{p_\perp}^2/\langle{p_\perp}^2\rangle}.
%$

%$With the gaussian expressions of $f_q(x,\bfk_\perp)$ and $D_q^h(z,\bfp _\perp)$

In this way the $\bf k_\perp$ integration in Eq. (\ref{sidis-Xsec-final}) can be 
performed analytically, leading to the result, valid up to ${O}(\kt/Q)$:
%$
%\frac{d^5\sigma^{\ell p \to \ell h X }}{d\xbj \, dQ^2 \, dz_h \, d^2\bfP _T} 
%\simeq 
%\sum_q \frac{2\pi\alpha^2e_q^2}{Q^4} \> f_q(\xbj) \> D_q^h(z_h) \biggl[ 
%(1+(1-y)^2)   - 4 \> 
%\frac{(2-y)\sqrt{1-y}\> \langle\kt^2\rangle \, z_h \, P_T}
%{\langle\pt^2\rangle \, Q}\> \cos \phi_h \biggr]
%\frac{1}{\pi\langle\pt^2\rangle} \, e^{-P_T^2/\langle\pt^2\rangle} 
%$
\bea
\frac{d^5\sigma^{\ell p \to \ell h X }}{d\xbj \, dQ^2 \, dz_h \, d^2{\bf P}_T} 
\simeq 
\sum_q \frac{2\pi\alpha^2e_q^2}{Q^4} \> f_q(\xbj) \> D_q^h(z_h) \biggl[ 
1+(1-y)^2  \nonumber \\ 
- 4 \> 
\frac{(2-y)\sqrt{1-y}\> \langle\kt^2\rangle \, z_h \, P_T}
{\langle\pt^2\rangle \, Q}\> \cos \phi_h \biggr]
\frac{1}{\pi\langle\pt^2\rangle} \, e^{-P_T^2/\langle\pt^2\rangle}\; , 
\eea
 where
$
\langle \pt^2 \rangle = \langle \ptq^2 \rangle + z_h^2 \langle \kt^2 \rangle\,.
$
The term proportional to $\cos \phi_h$ describes the Cahn effect~\cite{cahn}.

\begin{figure}[t]
%\hspace{-0.5cm}\parbox[l]{5.5cm}{\includegraphics[width=0.55\textwidth,bb = 10 370 540 670]
%{cahn.eps}}
\hspace{-3.5cm}\parbox[l]{5.5cm}{\includegraphics[width=0.4\textwidth]
{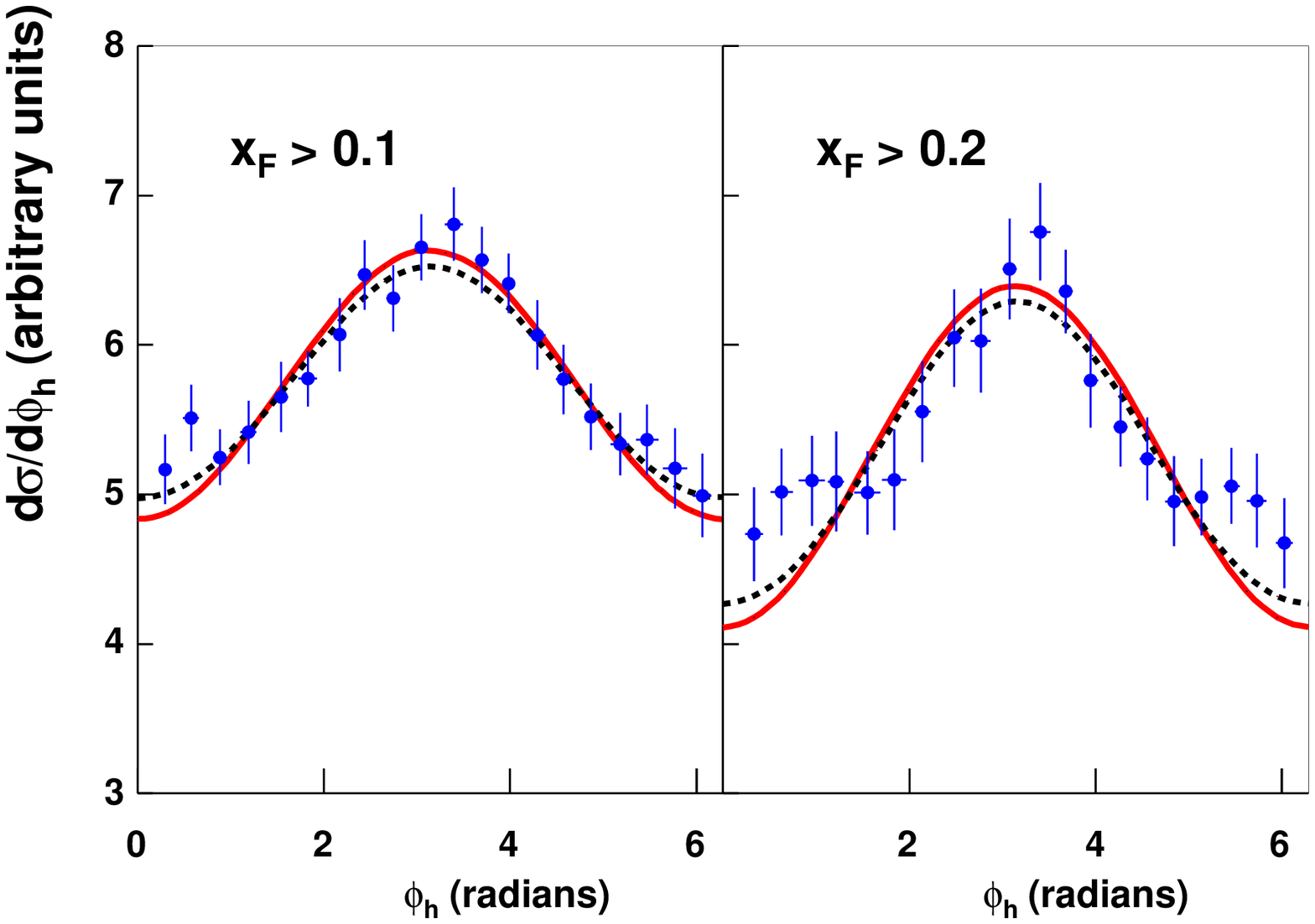}}\hspace{0.5cm}~\parbox[r]{5cm}{\includegraphics[width=0.55\textwidth,
,bb= 10 400 540 660]{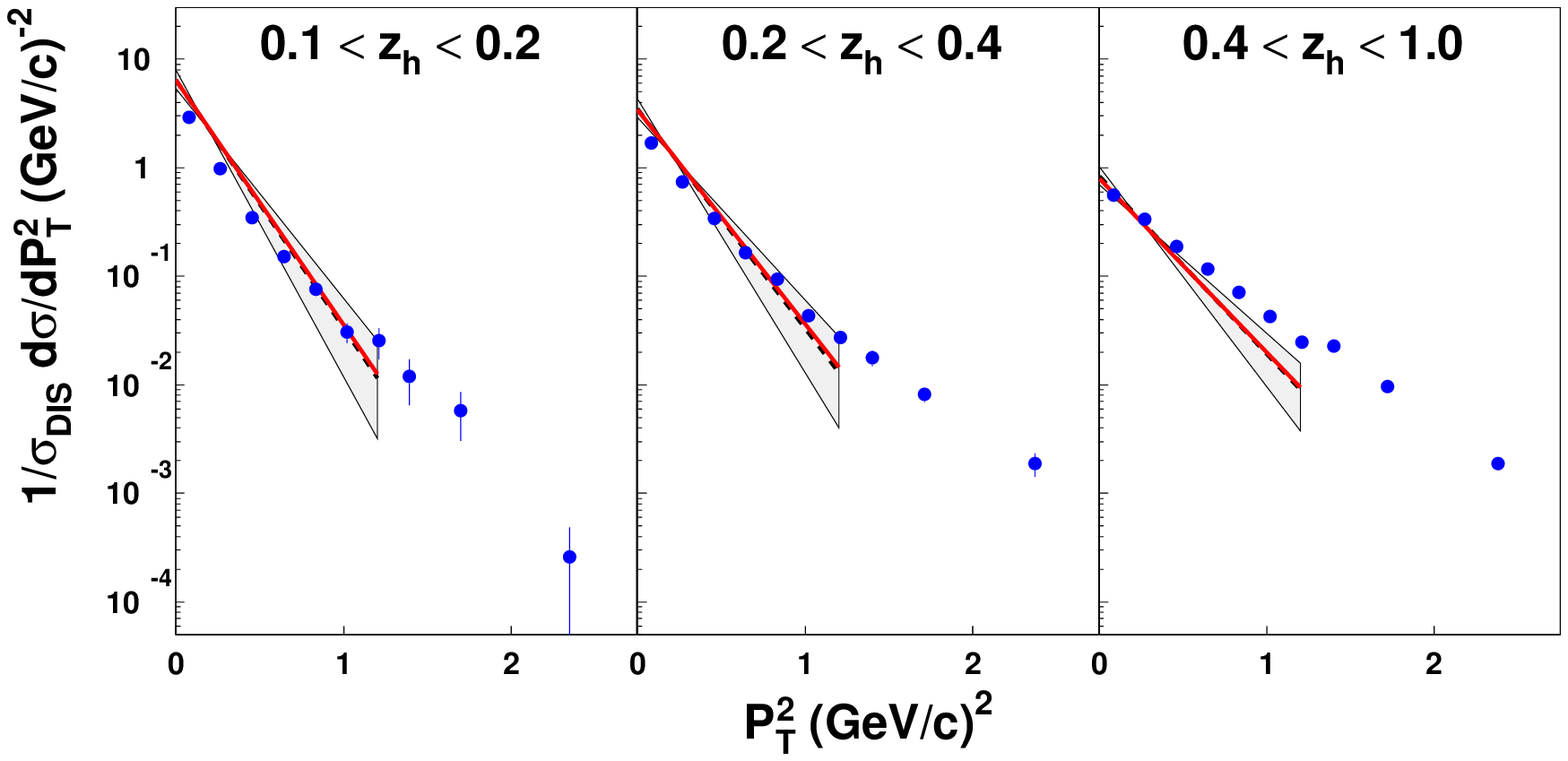}}
\vspace{-0.3cm}
\caption{
Description of the $\phi_h$ and $P_T$ dependence of the cross 
section.}
\label{fig:cahn}
\vspace{-0.4cm}
\end{figure}

By fitting the data~\cite{cahndata} on unpolarized 
SIDIS we obtain the following values of the parameters: 
$\langle\kt^2\rangle   = 0.25  \;{\rm (GeV/c)^2}
$, $\langle\ptq^2\rangle  = 0.20 \;{\rm (GeV/c)^2} \>.$ 
The results are shown in Fig.~\ref{fig:cahn}.

%%%%%%%%%%%%%%%%%%%%%%%%%%%%%%%%%%%%%%%%%%%%%%%%%%%%%%%%%%%%%%%%%%%%%%%%%%%
%%%%%%%%%%%%%%%%%%%%%%%%%%%%%%%%%%%%%%%%%%%%%%%%%%%%%%%%%%%%%%%%%%%%%%%%%%% 
%\section{Sivers effect in polarized SIDIS}
%\vspace{-1cm}
Such values are then used to compute the SSA for 
$\ell \, \pup \to \ell \, h \, X$ processes.
We considered the Sivers  mechanism~\cite{siv} alone.
The unpolarized quark (and gluon) distributions 
inside a transversely polarized proton 
%(generically denoted by $\pup$, with 
%$\pdown$ denoting the opposite polarization state) 
can be written as:
\be
f_ {q/\pup} (x,{\bf k_\perp}) = f_ {q/p} (x,\kt) +
\frac{1}{2} \, \Delta^N f_ {q/\pup}(x,\kt)  \;
{\bf S} \cdot (\hat {\bf P}  \times
\hat{\bf k}_\perp)\; , \label{poldf}
\ee
where ${\bf P}$ and $\bf S$ are respectively the proton momentum and transverse 
polarization vector, and $\bf k_\perp$ is the parton transverse momentum;
transverse refers to the proton flight direction. Eq.~(\ref{poldf}) leads to non
vanishing SSA, which can be calculated by substituting $f_ {q/p}$ by $f_ {q/\pup}$
in Eq.~(\ref{sidis-Xsec-final}). 

We parameterize, for each light quark flavour $q = u_v, d_v, u_s, d_s, \bar u, \bar d$, the 
Sivers function in the following factorized form:
$
\Delta^N f_ {q/\pup}(x,\kt) = 2 \, {\rm N}_q(x) \, h(\kt) \, 
f_ {q/p} (x,\kt)\; , $
where
${\rm N}_q(x) =  N_q \, x^{a_q}(1-x)^{b_q} \,
\frac{(a_q+b_q)^{(a_q+b_q)}}{a_q^{a_q} b_q^{b_q}}$ , 
$h(\kt) = \sqrt{2e} \, \frac{\kt} {M'} \, e^{-\kt^2/M'^{2}}$.

\begin{figure}[ht]
%\begin{center}
\includegraphics[width=0.5\textwidth,bb= 10 140 540 660]
{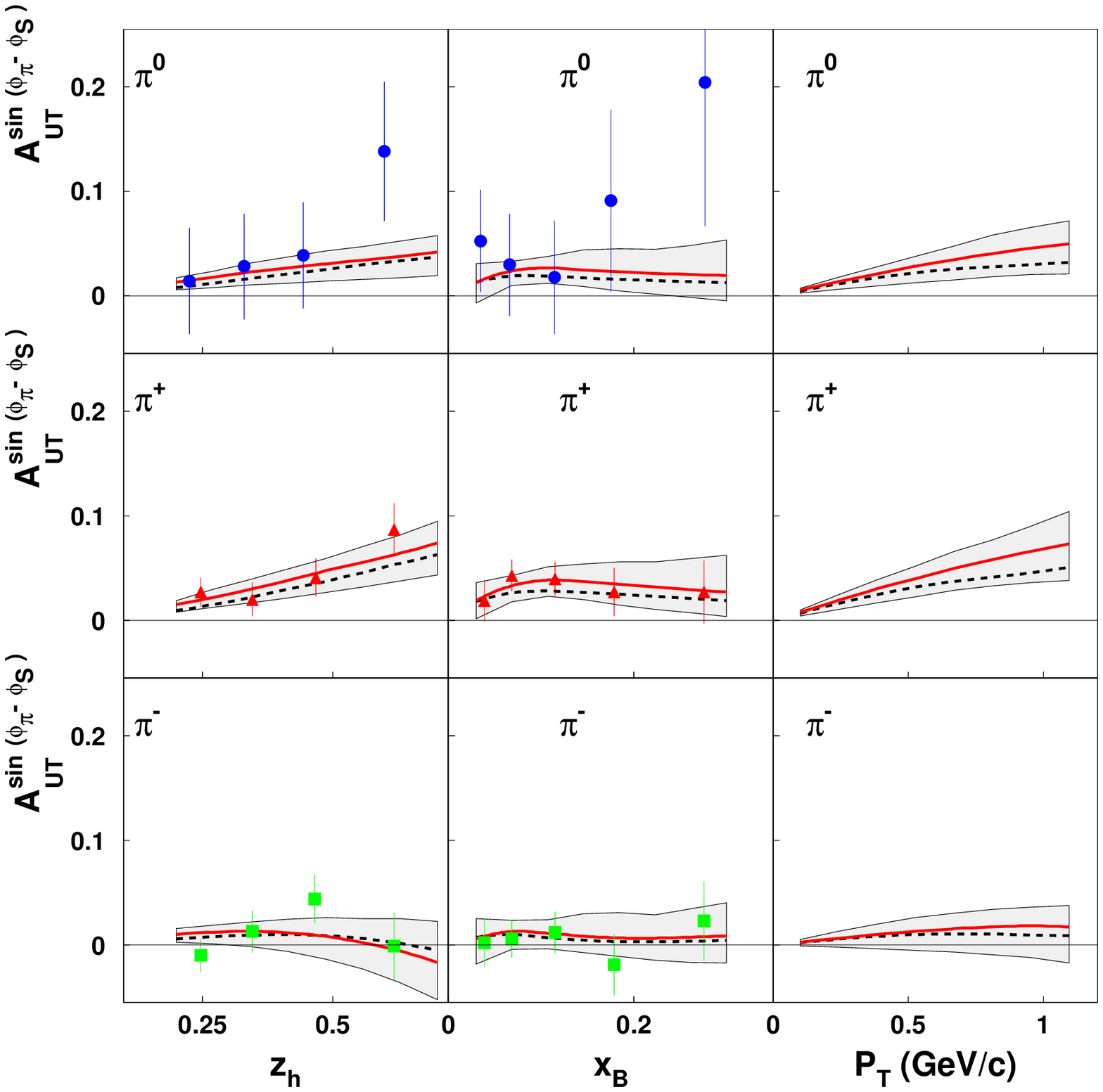}
\includegraphics[width=0.5\textwidth,bb= 10 140 540 660]
{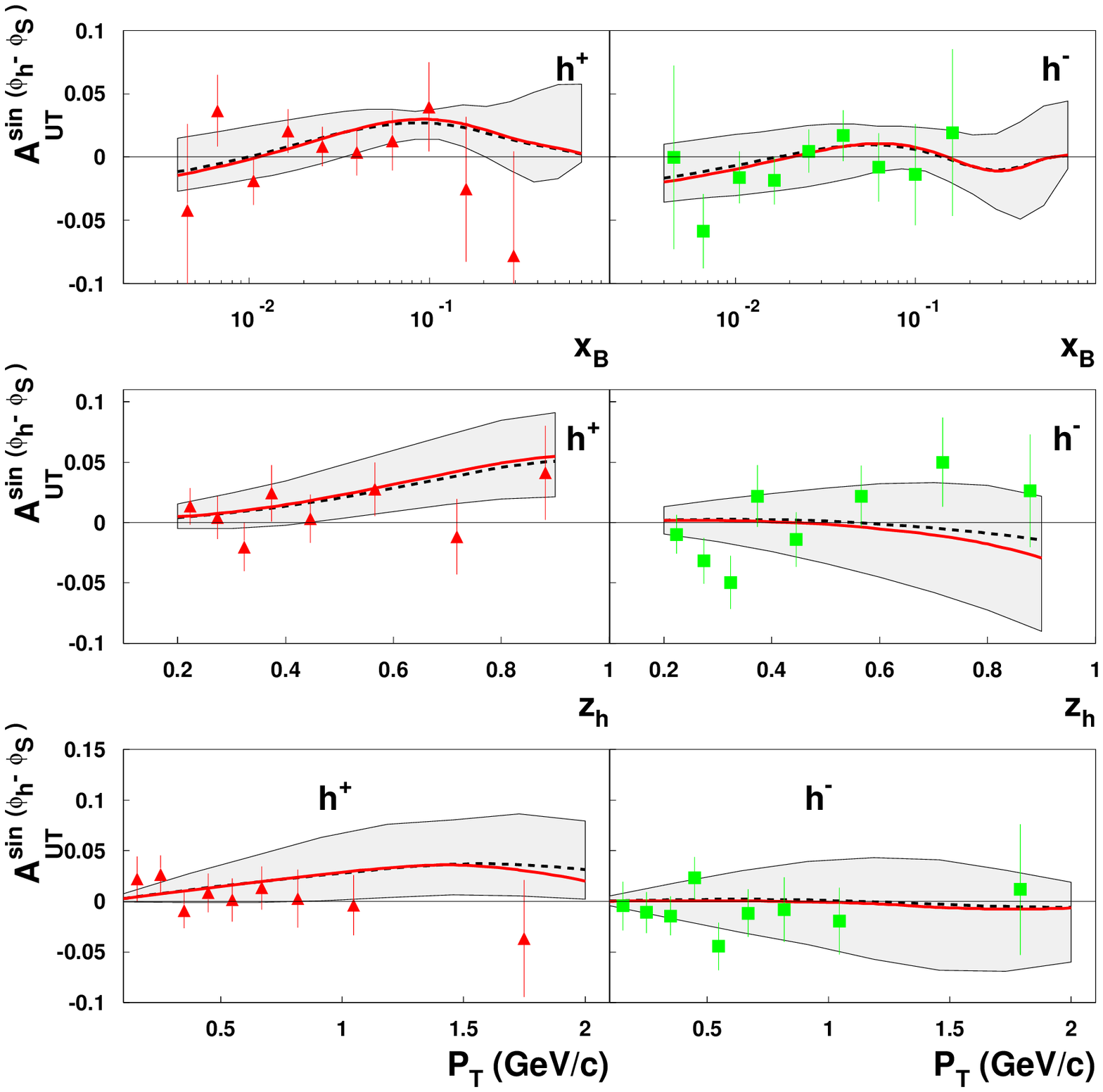}
\caption{\small HERMES~\cite{hermUT} (left) and COMPASS ~\cite{compUT} (right) data on $A_{UT}^{\sin(\phi_\pi-\phi_S)}$ for scattering
off a transversely polarized proton (deuterium) target and pion (hadron) production. The curves are
the results of our fit to the HERMES data and description of the COMPASS data , with exact kinematics (dashed line) or keeping only
terms up to ${O}({\kt}/{Q})$ (solid bold line). The shadowed region
corresponds to the theoretical uncertainty due to the parameter errors. 
}
\label{fig:aut}
\end{figure}

Our fit~\cite{our} to the HERMES data on $A_{UT}^{\sin(\phi_\pi-\phi_S)}$~\cite{hermUT} is presented in the left panel of Fig.~\ref{fig:aut}.

Having fixed all the parameters we can check the consistency of the model 
by computing $A_{UT}^{\sin(\phi_\pi-\phi_S)}$ 
 for charged hadron production in COMPASS
 experiment ~\cite{compUT}; our results are given in the right panel 
of Fig.~\ref{fig:aut}, showing a very good agreement with the data. 

%%%%%%%%%%%%%%%%%%%%%%%%%%%%%%%%%%%%%%%%%%%%%%%%%%%%%%%%%%%%%%%%%%%%%%%%%%%
%%%%%%%%%%%%%%%%%%%%%%%%%%%%%%%%%%%%%%%%%%%%%%%%%%%%%%%%%%%%%%%%%%%%%%%%%%%

%%%%%%%%%%%%%%%%%%%%%%%%%%%%%%%%%%%%%%%%%%%%%%%%%%%%%%%%%%%%%%%%%%%%%%%%%%%
\begin{figure}[ht]
%\begin{center}
\includegraphics[width=0.5\textwidth,height=0.23\textheight,bb= 10 140 540 660]
{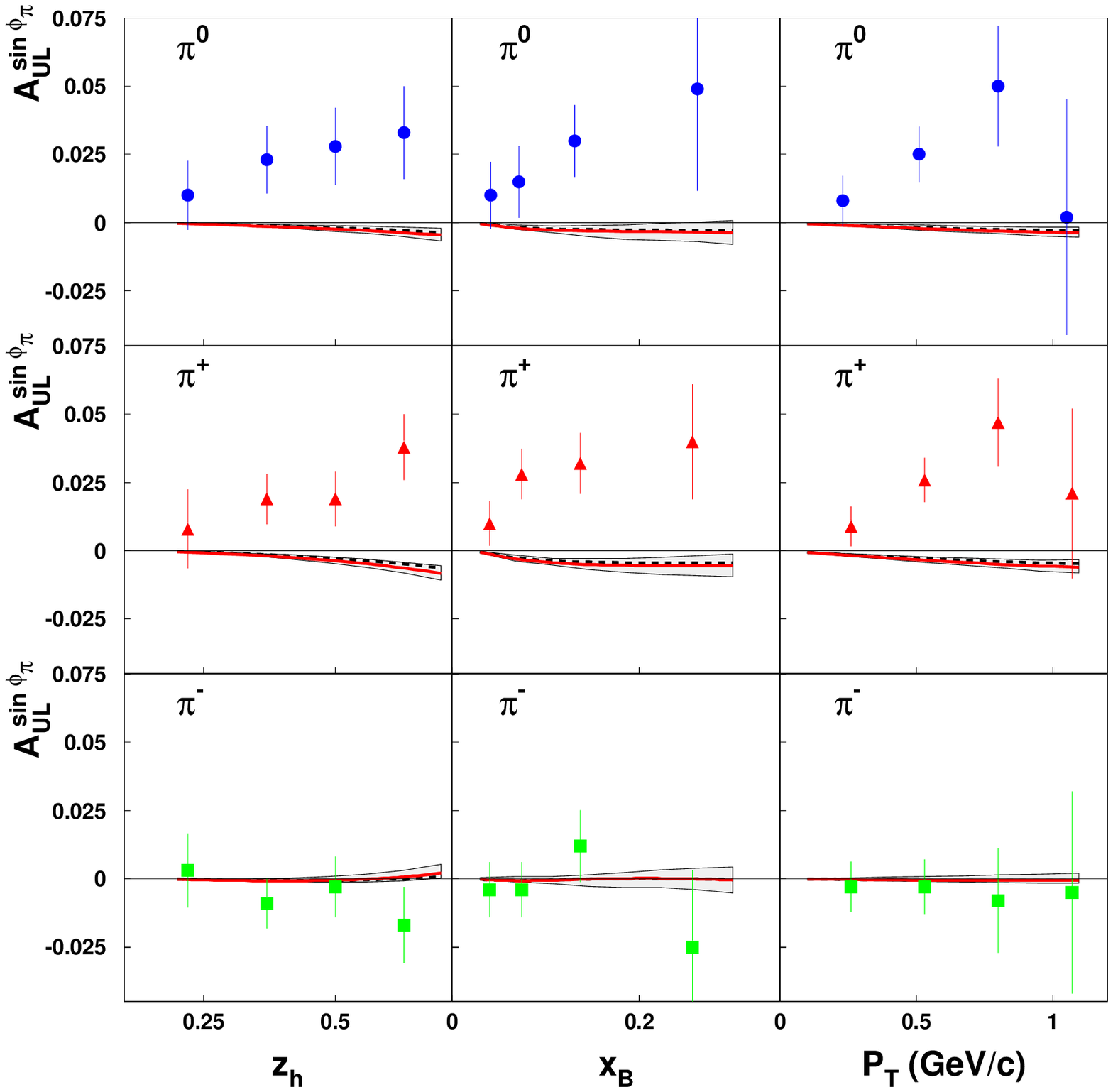}
\includegraphics[width=0.5\textwidth,height=0.23\textheight]
{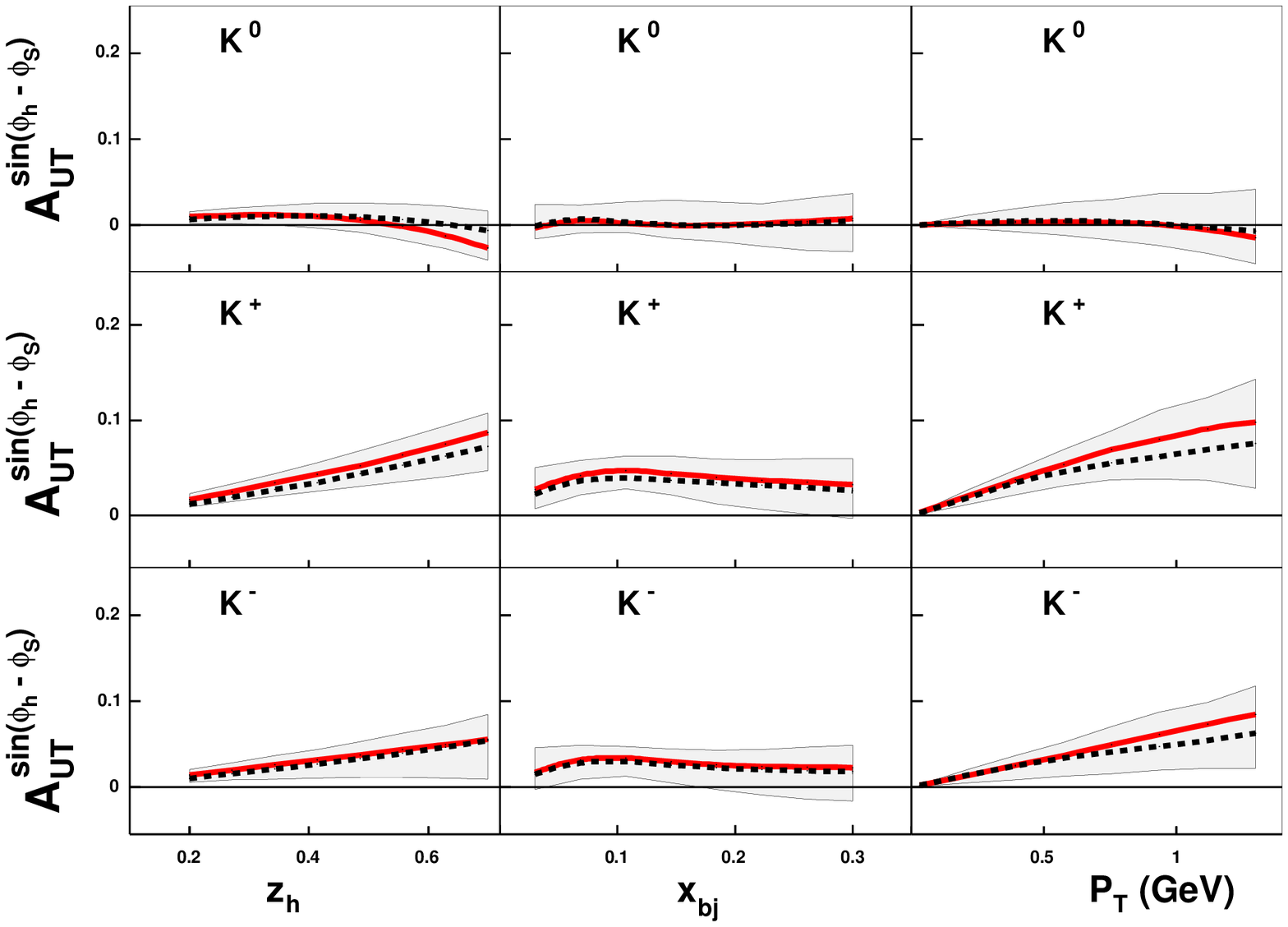}
\caption{\small HERMES data on $A_{UL}^{\sin(\phi_\pi)}$ ~\cite{hermUL} for scattering
off a longitudinally polarized proton target and pion production (left) and predictions of $A_{UT}^{\sin(\phi_K-\phi_S)}$ for kaon production at HERMES (right). }
\label{fig:autkaon}
\end{figure}
%%%%%%%%%%%%%%%%%%%%%%%%%%%%%%%%%%%%%%%%%%%%%%%%%%%%%%%%%%%%%%%%%%%%%%%%%%%

We also compute $A_{UT}^{\sin(\phi_K-\phi_S)}$ for kaon production,
which could be measured by HERMES. Our results are given in the right panel of  Fig. \ref{fig:autkaon}.

Finally, we consider the HERMES data on
$A_{UL}^{\sin\phi_\pi}$ obtained in the semi-inclusive
electro-production of pions on a longitudinally polarized hydrogen
target ~\cite{hermUL}. We have computed the Sivers contribution to
this quantity again with our
set of Sivers functions, and compared with data (see left panel of Fig. \ref{fig:autkaon}). Notice that no
agreement should be necessarily expected, as
$A_{UL}^{\sin\phi_\pi}$ can be originated also (even dominantly)
from the Collins mechanisms or higher-twist terms.

The HERMES data ~\cite{hermUT}
clearly show a non zero Sivers effect; by a comparison
with these data estimates of the Sivers functions for $u$ and
$d$ (both valence and sea) quarks have been obtained. These functions not
only describe well the HERMES data, but are also in agreement with
COMPASS preliminary data ~\cite{compUT}.

A phenomenological study of SSA and azimuthal dependences, within
a factorization scheme with unintegrated parton distribution and
fragmentation functions, is now possible. SIDIS processes with
measurements of the Cahn effect, and the various SSA
$A_{UL}^{\sin\phi_h}$, $A_{UT}^{\sin(\phi_h-\phi_S)}$ and
$A_{UT}^{\sin(\phi_h+\phi_S)}$ provide a rich ground to be further
explored, both theoretically and experimentally.

\end{document}